\documentstyle[twocolumn,prb,aps,floats,epsfig]{revtex}
\begin{document}
\twocolumn[\hsize\textwidth\columnwidth\hsize\csname 
@twocolumnfalse\endcsname
\title{Segregated tunneling-percolation model
for transport nonuniversality}
\author{C. Grimaldi$^1$, T. Maeder$^{1,2}$, P. Ryser$^1$, and S. Str\"assler$^{1,2}$} 
\address{$^1$ Institut de Production et Robotique, LPM,
Ecole Polytechnique F\'ed\'erale de Lausanne,
CH-1015 Lausanne, Switzerland}
\address{$^2$ Sensile Technologies SA, PSE, CH-1015 Lausanne, Switzerland}

\maketitle

\begin{abstract}
We propose a theory of the origin of transport nonuniversality in 
disordered insulating-conducting compounds based on the interplay between
microstructure and tunneling processes between metallic grains dispersed 
in the insulating host. We show that if the metallic phase is arranged
in quasi-one dimensional chains of conducting grains, then the distribution
function of the chain conductivities $g$ has a power-law divergence for 
$g\rightarrow 0$ leading to nonuniversal values of the transport critical exponent $t$.
We evaluate the critical exponent $t$ by Monte Carlo calculations on a
cubic lattice and show that our model
can describe universal as well nonuniversal behavior of transport depending 
on the value of few microstructural parameters. 
Such segregated tunneling-percolation model can describe the microstructure of
a quite vast class of materials known as thick-film resistors which display
universal or nonuniversal values of $t$ depending on the composition.

\end{abstract}
\vskip 2pc ]

\section{introduction}
\label{intro}

When the conductivity $\sigma$ of an insulating-conducting
compound is measured as a function of the volume concentration $p$
of the conducting phase, one finds that by reducing $p$ eventually
the system undergoes a conductor-to-insulator transition at a
particular critical value $p_c$ of the volume concentration. In
the critical region $0<p-p_c\ll 1$ the conductivity follows a
power law behavior of the form:
\begin{equation}
\label{eq1}
\sigma=\sigma_0(p-p_c)^t,
\end{equation}
where $\sigma_0$ is a prefactor which depends on the particular
system considered and $t$ is a positive number typically larger
than the unity.

Percolation theory explains the power law form of Eq.(\ref{eq1})
as being due to the lack of any cut-off length scale apart the
linear size of the sample and predicts that the exponent $t$ is
universal and depends only upon the dimensionality of the
system.\cite{stauffer} This prediction is confirmed by various
granular metals compounds and model systems which have been found
to follow Eq.(\ref{eq1}) with $t\simeq
2.0$,\cite{abeles,carcia1,lee,putten} that is the value obtained by
numerical calculations on three-dimensional random resistor
network (RRN) models.\cite{clerc}

In addition to systems showing universality, a large number of
disordered compounds displaying values of $t$ larger than
$t\simeq 2.0$ have been repeatedly
reported,\cite{pike,dejeu,carcia2,listki,balb,kusy,wu} so that in
the present situation it appears that $t$ can assume any value
between $t\simeq 2.0$ up to about $t\sim 6.0-7.0$.

Within percolation theory on a RRN, Kogut and Straley showed that
universality breakdown of transport may arise from anomalous
distributions of elemental conductivities.\cite{kogut} By
assigning to each neighbouring couple of sites on a regular
lattice a bond with finite conductivity $g$ with probability $p$
and zero conductivity with probability $1-p$, the resulting bond
conductivity distribution function becomes:
\begin{equation}
\label{distri1} \rho(g)=p h(g)+(1-p)\delta (g),
\end{equation}
where $\delta(g)$ is the Dirac delta-function and $h(g)$ is the
distribution function of the finite bond conductivities. For well
behaved $h(g)$, transport is universal and follows Eq.(\ref{eq1})
with $t=t_0\simeq 2.0$ for three dimensional lattices. Instead,
if $h(g)$ has a power law divergence for small $g$ of the form:
\begin{equation}
\label{distri2} 
\lim_{g\rightarrow 0} h(g) \propto g^{-\alpha},
\end{equation}
and $\alpha$ is larger than a critical value $\alpha_c$, 
Kogut and Straley showed that
transport is no longer universal and the conductivity exponent
becomes dependent on $\alpha$.\cite{kogut} Renormalization group analysis
predicts in fact that
\begin{equation}
\label{nonuni} 
t=\left\{
\begin{array}{ll}
t_0 & \mbox{if} \hspace{3mm} (D-2)\nu+\frac{1}{1-\alpha}< t_0 \\
(D-2)\nu+\frac{1}{1-\alpha} &  \mbox{if}\hspace{3mm}
(D-2)\nu+\frac{1}{1-\alpha}> t_0
\end{array}
\right.,
\end{equation}
where $D$ is the dimensionality of the lattice and $\nu$  is the
correlation-length exponent ($\nu=4/3$ for $D=2$ and $\nu\simeq
0.88$ for $D=3$).\cite{straley,machta,alava} For $D=3$ and by using $t_0\simeq 2.0$ 
and $\nu\simeq 0.88$ the critical value of the exponent 
is $\alpha_c\simeq 0.107$.

Microscopic models which may justify Eq.(\ref{distri2}) are the
random void (RV) model proposed by Halperin, Seng and
Fen,\cite{halperin} and the tunneling-percolating model of
Balberg.\cite{balb} The RV model describes a system of insulating
spheres (or disks in two dimensions) embedded randomly in a
continuous conducting material. In this situation, transport is
dominated by the conductivity of the narrow necks bounded
by three interpenetrating insulating spheres. Such necks have a wide distribution in 
widths resulting in a wide distribution of conductivities.
The original
formulation of the RV model predicted $t=t_0+0.5$ for the
conductivity exponent of the whole sample. A recent
generalization  of the RV model by Balberg has shown that $t$ can
assume even higher values and that in principle is not bounded
above.\cite{balbRV}

In the tunneling-percolating model of Ref.[\onlinecite{balb}], transport
is assumed to be dominated by quantum tunneling between
neighbouring conducting particles dispersed in an insulating
medium. If the distribution function $P(r)$ of the distance $r$
between two neighbouring particles decays with $r$ much slower
than the tunneling decay $\exp(-2r/\xi)$, where $\xi$ is the
localization length, then the tunneling conductivity distribution
function $h(g)$ can be shown to behave as Eq.(\ref{distri2}) with
$\alpha\simeq 1-\xi/2a$, so that the transport exponent $t$
becomes dependent of the mean tunneling distance $a$.\cite{balb}
Interactions between the conducting and insulating phases as well
as properties of the microstructure are argued to concur to the
$r$-dependence of $P(r)$. Due to the complexity of the problem,
explicit calculations of the interparticle distance distribution
function are missing and one must relay on phenomenological forms
of $P(r)$.

In this paper we provide a microscopic derivation of $P(r)$ which
has been inspired by
the peculiar microstructure observed in a particular class of
insulating-conducting compounds: the so-called thick-film
resistors (TFRs). These compounds are based on RuO$_2$ (or
Bi$_2$Ru$_2$O$_7$, Pb$_2$Ru$_2$O$_6$, and IrO$_2$) grains mixed
and fired with glass powders.\cite{prude1} Typically, TFRs are
often in a segregated structure regime in which large regions of
glass constraint the much smaller
conducting grains to be segregated in between the interstices of
neighbouring glass grains. Micrographs reveal that the conducting
grains are arranged in a network of filaments spanning the entire
sample.\cite{pike,chiang,hrovat} By taking into account the quasi-one dimensional
structure of such filaments and by neglecting interactions with
the insulating phase, we show that the resulting $P(r)$ can decay
much slower than the tunneling decay leading to nonuniversal
behavior of transport.

This paper is organized as follows. In the next section we
construct a RRN model which captures the essential structure of
the filamentary network of TFRs and calculate the resulting $P(r)$
and the distribution function of the conductivity of filaments.
In Sec.\ref{results} we perform Monte Carlo calculations and
calculate the conductivity exponent $t$ for a variety of
situations. The last section is devoted to discussions and
conclusions.

\section{the model}
\label{model}
Before describing our model in details, we find it useful to first
discuss in general the interplay between the spatial distribution
of the conducting phase within the insulating matrix and transport properties.
Let us consider a generic insulating-conducting compound where
the conducting grains are embedded in an insulating host. In this situation,
electron transfer is governed by electron tunneling
from grain to grain. The grain charging energy and the Coulomb
interaction between charged grains affects the overall transport properties
especially regarding their behavior in temperature. Here we focus on
systems where the temperature is high enough to possibly neglect
charging and Coulomb effects so that the main electron transfer 
is dominated solely by tunneling leading to intergrain conductivity
of the form:
\begin{equation}
\label{tunnel1}
\sigma(r)=\sigma_0 e^{-2(r-\Phi)/\xi},
\end{equation}
where $\sigma_0$ is a constant which can be set equal to the unity without loss of
generality, $\xi \propto 1/\sqrt{V}$ is the tunneling factor (or localization length)
and $V$ is the intergrain barrier potential. 
In Eq.(\ref{tunnel1}) we have approximated the
conducting grains by spheres of diameter $\Phi$ and $r$ is the distance between
the centers of two spheres which we treat as impenetrable ($r\ge \Phi$). 

Due to the exponential decay of Eq.(\ref{tunnel1}), contributions to $\sigma(r)$
from far away spheres can be neglected,\cite{balb,toker} so that from now on $r$ denotes
the distance of two nearest-neighbouring spheres. Hence, the ensemble
dependence of $\sigma(r)$
upon $r$ is completely defined by the distribution function $P(r)$ of the
distance between nearest-neighbouring spheres. In fact, once $P(r)$ is known, 
the conductivity distribution function $h(g)$ can be obtained as follows:
\begin{equation}
\label{distria}
h(g)=\int \!dr P(r)\delta[g-\sigma(r)].
\end{equation}
In this preliminary discussion, we are interested in studying how the form of $P(r)$
affects $h(g)$ via Eq.(\ref{distria}) and which are the requisites of $P(r)$
which eventually could generate a power-law distribution function 
as that of Eq.(\ref{distri2}). 
As already pointed out, $P(r)$ depends on the microstructure
of the composite and on eventual interactions between the insulating and the 
conducting phases. In principle, therefore, the form of $P(r)$ depends on the 
particular composite considered. However, if we imagine that interactions can be neglected, 
then it is natural to assume that the conducting spheres are Poisson
distributed within the insulating phase. Then if $D$ is the dimensionality of the system,
by following Refs.[\onlinecite{macdonald,torquato}]
the nearest-neighbour distance distribution function is approximatively
of the form:
\begin{equation}
\label{pappr}
P(r)\sim \frac{e^{-(r/a_D)^D}}{a_D},
\end{equation}
where $a_D$ is a constant depending on the mean distance
between neighbouring spheres. Equation (\ref{pappr}) is an asymptotic approximation of the
true $P(r)$ and is valid only in the $r/a_D\gg 1$ limit. This is however 
the limiting region of interest to us since it governs, via Eq.(\ref{tunnel1}), 
the $g\ll 1$ regime.
It is also worth to point out that
Eq.(\ref{pappr}) holds true for penetrable as well as impenetrable (hard-core) spheres,
the only difference being in the explicit expression for $a_D$ which is however of not
importance at the moment.\cite{torquato}

By inserting Eqs.(\ref{tunnel1},\ref{pappr}) into Eq.(\ref{distria}), the resulting conductivity
distribution function becomes:
\begin{eqnarray}
\label{apprb}
h(g)&\sim & \int \!\frac{dr}{a_D}e^{-(r/a_D)^D}\delta[g-\sigma(r)] \nonumber \\
&=& \frac{\xi}{2a_D}\frac{1}{g}\exp\left(-\frac{\xi}{2a_D}\ln g^{-1}\right)^D,
\end{eqnarray}
which after some manipulations reduces to:
\begin{equation}
\label{apprc}
h(g)\sim \frac{\xi}{2a_D} g^{(\frac{\xi}{2a_D})^D(\ln g^{-1})^{D-1}-1}.
\end{equation}
For $D=2$ and $D=3$, the $g\rightarrow 0$ limit of the above expression goes to zero 
irrespectively of the value of $\xi/2 a_D$. In this case therefore no power-law
divergence of $h(g)$ is encountered and, as discussed in the introduction, transport
is governed by the universal critical exponent $t=t_0\simeq 2.0$.
Instead, when $D=1$, equation (\ref{apprc}) becomes:
\begin{equation}
\label{apprd}
h(g) \sim \frac{\xi}{2a_1}g^{\frac{\xi}{2a_1}-1},
\end{equation}
which is exactly of the form of Eq.(\ref{distri2}) if we identify $\alpha$ with
$1-\xi/2a_1$. We have arrived therefore at the result that if the spheres are 
Poisson distributed along a one-dimensional line, the resulting conductivity
distribution function has a power-law behavior for small $g$ and, consequently,
transport is nonuniversal for sufficiently large values of $1-\xi/2a_1$.

The difference between the $D=2,3$ and the $D=1$ cases stems from the decay of 
Eq.(\ref{pappr}) which for $D=2,3$ is much too fast with respect to the simple 
exponential decay of Eq.(\ref{tunnel1}). In fact from Eq.(\ref{distria}) 
it is simple to show that as long as $\lim_{r\rightarrow \infty} P(r)/\sigma(r)=0$ 
then $\lim_{g\rightarrow 0} h(g)=0$ irrespectively of the detailed structure of $P(r)$.
Hence to construct a RRN model having $h(g)$ of the form of Eq.(\ref{distri2})
we must consider forms of $P(r)$ whose decay for $r\rightarrow \infty$ is sufficiently
slow. The result of Eq.(\ref{apprd}) suggests that for this scope one-dimensionality
is an important ingredient, at least as long as interactions between conducting
and insulating phases can be neglected. 

Among the various insulating-conducting compounds,
thick film resistors are systems whose microstructure can be appropriately described
in terms of quasi one-dimensional units.
Let us consider the highly non-homogeneous microstructure typical
of TFRs. These systems are constituted by a mixture of large
glassy particles (with size $L$ of order $1$-$3$ $\mu$m) and
small conducting grains of size $\Phi$ typically varying between
$\sim 10$ nm up to $\sim 200$ nm. Due to the high values of
$L/\Phi$, the small metallic grains tend to occupy the narrow
regions between the much larger insulating zones leading to a
filamentary distribution of the conducting phase.\cite{pike,chiang,hrovat} 
A classical
model to describe such a segregation effect was proposed already
in the 1970's by Pike.\cite{pike} This model replaces the glassy
particles by insulating cubes of size $L\gg \Phi$ whose edges can
be occupied by chains formed by adjacent metallic spheres of
diameter $\Phi$. Let us assume that an edge has probability $p$
of being occupied by a chain of $n+1$ spheres and probability $1-p$
of being empty. As depicted in Fig.1, the set of occupied and
empty edges form a cubic lattice spanning the entire sample.

\begin{figure}[t]
\protect
\centerline{\epsfig{figure=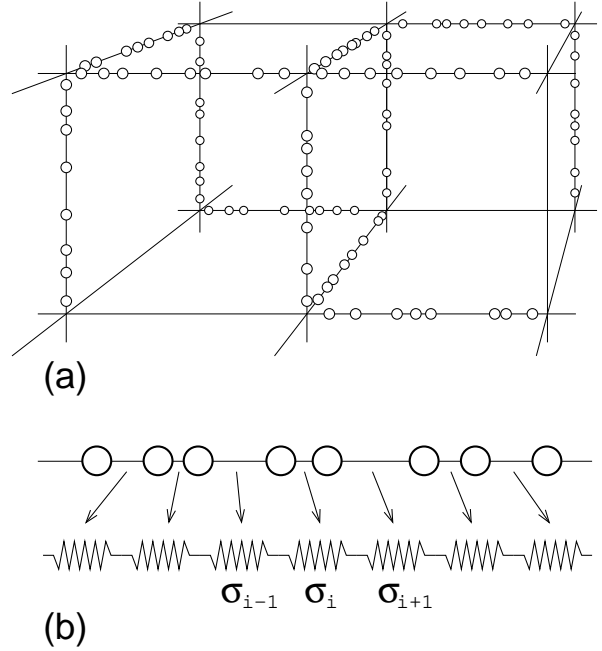,width=20pc,clip=}}
\caption{(a): pictorial representation of the segregated tunneling-percolation
model. The cubes represent insulating grains while the spheres are conducting
particles. The spheres are arranged to occupy the edges of the insulating cubes
with probability $p$. The total ensemble of occupied an unoccupied edges forms
a cubic lattice spanning the entire sample. (b): equivalence between 
an edge occupied by $n+1$ spheres and a conducting element. The set of inter-spheres
tunneling conductivities is equivalent to a conductor with $n$ resistive
elements with conductivities $\sigma_i$ in
series. The fluctuation in distance between two neighbouring spheres
leads to fluctuating tunneling conductivities. With this equivalence, the
model depicted in (a) can be considered as a bond-percolation model 
where a fraction $p$ of bonds has variable conductivities and a 
fraction $1-p$ is insulating.}
\label{fig1}
\end{figure}

\begin{figure}[t]
\protect
\centerline{\epsfig{figure=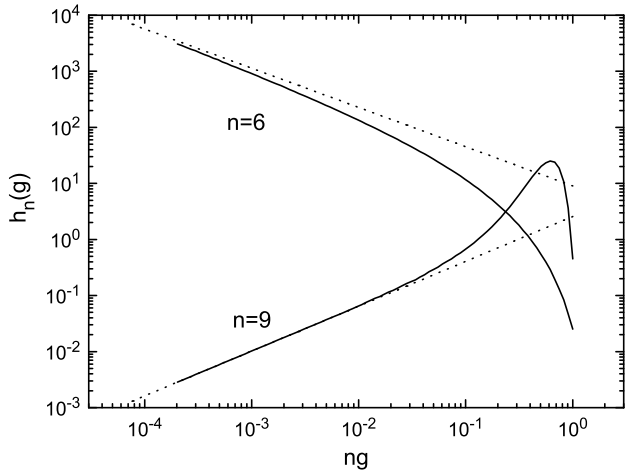,width=22pc,clip=}}
\caption{Distribution function $h_n(g)$ of the conductivity of the occupied channels
for $L/\Phi=10$, $\xi/\Phi=0.2$ and different values of $n$.
Solid lines are the result of a numerical calculation of Eq.(\ref{distri4}) while the 
dotted lines are the asymptotic results of Eq.(\ref{appa3}).}
\label{fig2}
\end{figure}

To define the RRN relevant for this model we proceed as follows.
The conductivity $g$ of a single occupied channel is governed
by the conductivities of the metallic spheres and those between
pairs of two-neighbouring spheres. The conductivity of the metallic spheres
adds only a negligible contribution to $g$ which is then given by $n$ 
conductivities $\sigma_i$ of pairs of nearest-neighbouring spheres in series:
\begin{equation}
\label{sigma}
g^{-1}=\sum_{i=1}^n \frac{1}{\sigma_i}.
\end{equation} 
We assume that the
inter-sphere conductivities $\sigma_i$ are due to tunneling processes
between two adjacent spheres so that their sphere-to-sphere distance $r$
dependence is that of Eq.(\ref{tunnel1}). For TFRs, the tunneling hypothesis is well
sustained by their high values of piezoresistance ({\it i.e.}, the
strain sensitivity of transport),\cite{prude1} and the low
temperature dependence of transport indicating some kind of
assisted hopping. As done in the introductory part of this section, we neglect
interactions between the
insulating and conducting phases, and assume that the sphere centers are
Poisson distributed along the cube edge. In doing so, we implicitly assume
that finite size effects of the channels can be neglected and that 
periodic boundary conditions are applied. In this way the last sphere on one end
of the channel is identified with the first one on the opposite end, so that 
we have $n$ individual spheres and $n$ inter-sphere tunneling junctions.
In this situation, 
the distances $r$ change according to the
distribution function $P_n(r)$ of the nearest-neighbour distances $r$ of
$n$ impenetrable spheres
arranged randomly in a quasi one dimensional channel. By
following Ref.\cite{torquato}, $P_n(r)$ can be calculated exactly
and it is given by:
\begin{equation}
\label{distri5}
P_n(r)=\frac{1}{a_n-\Phi}e^{-(r-\Phi)/(a_n-\Phi)}\Theta(r-\Phi),
\end{equation}
where $\Theta$ is the step function and
\begin{equation}
\label{tunnel2} 
a_n=\frac{\Phi}{2}\left(1+\frac{L}{n\Phi}\right),
\end{equation}
is the mean inter-sphere (center-to-center) distance. In the above
expression $n\Phi/L$ cannot be larger than the unity since no more
than $L/\Phi$ spheres can be accommodated inside a channel. Note that the asymptotic
expression Eq.(\ref{pappr}) for $D=1$ coincides with Eq.(\ref{distri5}) if $a_1$
is identified with $a_n-\Phi$. Hence the
distribution function $f(\sigma)$ of the inter-sphere conductivities 
should be of the same form of Eq.(\ref{apprd}). In fact:
\begin{equation}
\label{distri6} 
f(\sigma)=\int_0^{\infty} \!dr\,
P_n(r)\,\delta[\sigma-\sigma(r)] =(1-\alpha_n)\sigma^{-\alpha_n},
\end{equation}
where
\begin{equation}
\label{alfa} 
\alpha_n=1-\frac{\xi/2}{a_n-\Phi}.
\end{equation}
Having obtained an explicit expression for the distribution
function $f(\sigma)$ of the inter-spheres conductivities, we can now
calculate the total distribution function $h_n(g)$ of the whole
channel. From Eq.(\ref{sigma}), $h_n(g)$ can be defined as:
\begin{equation}
\label{distri4} 
h_n(g)\!=\!\!\int\! \!d\sigma_1\ldots d\sigma_n
f(\sigma_1)\ldots f(\sigma_n) \nonumber \\
\delta\!\!\left[g-
\left(\sum_{i=1}^{n}\frac{1}{\sigma_i}\right)^{\!-1}\right],
\end{equation}
which, by using Eq.(\ref{distri6}), reduces to:
\begin{eqnarray}
\label{appa1}
h_n(g)&=&(1-\alpha_n)^n\int
d\sigma_1\ldots d\sigma_n\left(\prod_{i=1}^n\sigma_i\right)^{-\alpha_n}\nonumber \\
&\times&\delta\!\left[g-
\left(\sum_{i=1}^{n}\frac{1}{\sigma_i}\right)^{-1}\right]\nonumber \\
&=&(1-\alpha_n)^n g^{-\alpha_n}\int
d\sigma_1\ldots d\sigma_n\left(\sum_{i=1}^{n}\prod_{j\neq i}^n\sigma_j\right)^{-\alpha_n}\nonumber \\
&\times&\delta\!\left[g-
\left(\sum_{i=1}^{n}\frac{1}{\sigma_i}\right)^{-1}\right].
\end{eqnarray}
It is clear that $h_n(g)$ behaves as $g^{-\alpha_n}$
for $g\ll 1$ since the integral appearing in the last equality of the above
expression is well behaved in the $g\rightarrow 0$ limit.
In fact the $g\rightarrow 0$ limit of the Dirac $\delta$-function appearing
in Eq.(\ref{appa1}) reduces to:
\begin{equation}
\label{appa2} 
\lim_{g\rightarrow 0}\delta\!\left[g-
\left(\sum_{i=1}^{n}\frac{1}{\sigma_i}\right)^{-1}\right]=
\sum_{l=1}^{n}\delta(\sigma_l)\left(\frac{\sum_{i=1}^{n}\prod_{j\neq
i}\sigma_j}{\prod_{i\neq l}\sigma_i}\right)^2, \nonumber\\
\end{equation}
so that, finally:
\begin{eqnarray}
\label{appa3} 
h_n(g)&\simeq &(1-\alpha_n)^n g^{-\alpha_n}\!\!
\int \! \!d\sigma_1\ldots d\sigma_n\!\sum_{l=1}^{n}
\delta\!(\sigma_l)\!\left(\!\prod_{i\neq l}\sigma_i\!\right)^{\!-\!\alpha_n}\nonumber \\
&=&n(1-\alpha_n)g^{-\alpha_n} \,\,\,\, \,\,\,{\rm for}\,\,\, g\ll 1.
\end{eqnarray}
The above equation is the main result of this paper, {\it i. e.},
the distribution function of the occupied channel conductivities $h_n(g)$
is of the same form of Eq.(\ref{distri2}). 
In this situation, for sufficiently large values of $\alpha_n$ the RRN
conductivity can behave in a nonuniversal way with exponent $t > 2.0$.
The condition for universality breakdown is given by Eq.(\ref{nonuni})
which for a three-dimensional network implies $\alpha_n>\alpha_c\simeq 0.107$.
From Eqs.(\ref{tunnel2},\ref{alfa}) this condition corresponds to:
\begin{equation}
\label{nc}
n < n_c=\frac{1-\alpha_c}{1-\alpha_c+\xi/\Phi}L/\Phi,
\end{equation}
so that, for fixed values of $\xi/\Phi$ and $L/\Phi$, the value of the 
transport exponent $t$ is governed solely by the number of spheres that 
can be arranged within the occupied one-dimensional channels. The overall behavior
of $h_n(g)$ is reported in Fig.\ref{fig2} where we report a numerical calculation
of Eq.(\ref{distri4}) (solid lines) together with the asymptotic behavior 
obtained in Eq.(\ref{appa3}) (dotted lines). In this example we have set
$\xi=2$ nm, $\Phi=10$ nm and $L=0.1$ $\mu$m corresponding to $L/\Phi=10$,
$\xi/\Phi=0.2$ and $n_c\simeq 8.17$. For $n = 9 > n_c$ the distribution 
function goes to zero as Eq.(\ref{appa3}) with $\alpha_n\simeq -0.244$ while
for $n = 6 < n_c$ $h_n(g)$ diverges for $g\rightarrow 0$ with exponent
$\alpha_n=0.7$. Since $\alpha_c\simeq 0.107$, we expect that for $n=9$ transport
is universal while for $n=6$ the exponent $t$ becomes larger than $t_0\simeq 2.0$
as in Eq.(\ref{nonuni}).

\begin{figure}[t]
\protect
\centerline{\epsfig{figure=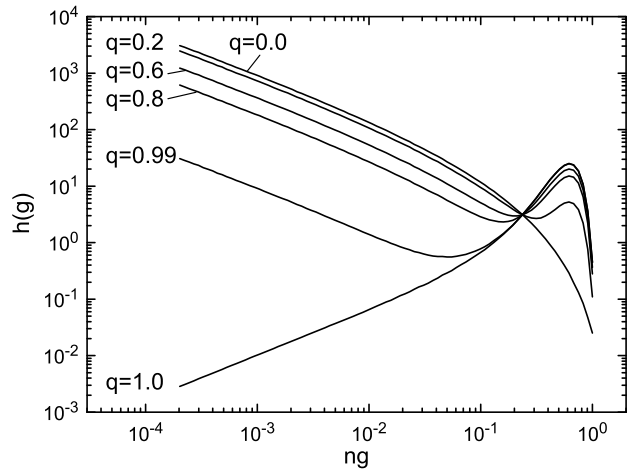,width=22pc,clip=}}
\caption{Channel conductivity distribution function $h(g)$, Eq.(\ref{distri3}), for the
bimodal distribution of Eq.(\ref{bimodal}) and for different values of $q$.
$L/\Phi=10$, $\xi/\Phi=0.2$, $n1=9$ and $n2=6$.}
\label{fig3}
\end{figure}

Before discussing our numerical results on the RRN conductivity, it is worth
to point out that our model can be easily generalized to consider also situations
in which the number of spheres accommodated in the one-dimensional channels 
is not fixed. More specifically, if ${\mathcal P}_{n'}$ is the distribution function
of the number $n'$ of spheres, then the distribution function of the occupied channels
is generalized to
\begin{equation}
\label{distri3}
h(g)=\sum_{n'}{\mathcal P}_{n'} h_{n'}(g).
\end{equation}
As an instructive case let us consider a bimodal distribution of the form:
\begin{equation}
\label{bimodal}
{\mathcal P}_{n'}=q\delta_{n'\!,n1}+(1-q)\delta_{n'\!,n2},
\end{equation}
where $0\leq q\leq 1$. For $q=0$ or $q=1$ we recover the previous case in which 
the occupied channels have the same number of spheres and whether transport is
universal or not depends on the specific values $n1$ (for $q=1$) or $n2$ (for $q=1$).
An interesting case is given by $0< q < 1$ and $n2 < n_c < n1$ according to which
there is a concentration $q$ of channels conductivities with distribution function
with exponent $\alpha_{n1} < \alpha_c$ and $1-q$ channels with $\alpha_{n2}> \alpha_c$.
This case is depicted in Fig.\ref{fig3} where $n1=9$ and $n2=6$ and, as in 
Fig. \ref{fig2}, $L/\Phi=10$ and $\xi/\Phi=0.2$. For all values $q < 1$ the 
$g\rightarrow 0$ limit is governed by the diverging part of the total distribution 
function. In this case we expect that at the critical point the transport 
exponent in not universal for any value $q < 1$. Note however that for $q$ sufficiently
close to the unity, the asymptotic regime is reached for relatively small values
of the conductivity. As we shall see in the next section, this has the effect of
shrinking the region where criticality sets in with $t=\nu+1/(1-\alpha_{n6})$.

\section{Monte Carlo results on the cubic lattice}
\label{results}
In this section we discuss our Monte Carlo calculations
for the conductivity $\sigma$ of the RRN model defined in the last section.
In constructing the RRN we must first implement numerically the conductivity
of the channels occupied by a given number $n$ of spheres. If $x_i$ ($i=1,\ldots,n$)
is a set of random numbers equally distributed in the interval $(0,1)$ then
it is easily found that the channel conductivity $g$ having Eq.(\ref{appa1}) 
as its corresponding distribution function is:
\begin{equation}
\label{num1}
g=\left[\sum_{i=1}^n x_i^{1/(\alpha_n -1)}\right]^{-1}.
\end{equation}
The RRN is then defined to have a fraction $p$ of channels (in a cubic lattice)
with $g$ as given by Eq.(\ref{num1}) and a fraction $1-p$ with $g=0$.
The generalization to a bimodal distribution, Eqs.(\ref{distri3},\ref{bimodal}), is
straightforward.

To calculate numerically the transport exponent $t$ we use the transfer-matrix 
method of Derrida and Vannimenus applied to a simple cubic lattice of $N-1$ sites in
the $z$ direction,
$N$ sites along $y$ and ${\mathcal L}$ along the $x$ direction.\cite{derrida1} 
Periodic boundary 
conditions are used  in the $y$-direction while to the top plane is applied
a unitary voltage and the bottom plane is grounded to zero.\cite{derrida2,normand} 
For ${\mathcal L}$
sufficiently large (${\mathcal L}\gg N$) this method permits to calculate the conductivity
per unit length of a cubic lattice. We calculate the conductivity $\sigma_N$
for different
linear sizes $N$ at the percolation thresold $p_c\simeq 0.2488126$ for bond
percolation on a cubic lattice,\cite{ziff} and then we extract by least-square fits
the critical exponent $t$ from the finite size scaling relation:\cite{normand,lobb1}
\begin{equation}
\label{scal}
\sigma_N=a N^{-t/\nu}(1+b N^{-\omega}),
\end{equation}
where $\nu\simeq 0.88$ is the correlation length exponent, $a$ and $b$ are
constants and $\omega$ is the first correction to the scaling exponent $t/\nu$.
In performing the calculations we have considered the following geometries:
$N=6$ (${\mathcal L}=5\times 10^7$), $N=8$ (${\mathcal L}=2\times 10^7$), 
$N=10$ (${\mathcal L}=1\times 10^7$),
$N=12$ (${\mathcal L}=8\times 10^6$), $N=14$ (${\mathcal L}=2\times 10^6$), 
and $N=16$ (${\mathcal L}=2\times 10^6$).

\begin{figure}[t]
\protect
\centerline{\epsfig{figure=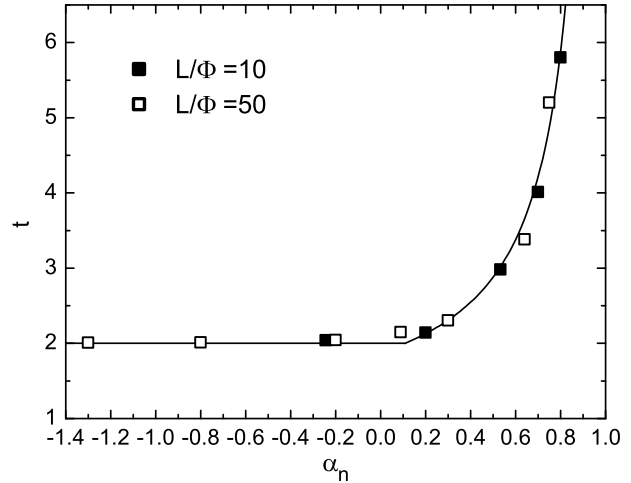,width=22pc,clip=}}
\caption{Critical exponent $t$ as a function of the tunneling exponent
$\alpha_n$ for $\xi/\Phi=0.2$ and different values of $L/\Phi$ and of the number $n$
of inter-sphere tunneling junctions accommodated within the occupied channels 
of a cubic random-resistor network.
From left to right: $n=9,\,8,\ldots,5$ for $L/\Phi=10$ (filled squares) 
and $n=46,\, 45,\, 43,\, 41,\, 39,\, 33,\, 27$ for $L/\Phi=50$ (open squares).
The solid curve is the theoretical result $t=t_0\simeq 2.0$ for 
$\alpha_n < \alpha_c\simeq 0.107$ and $t=\nu+1/(1-\alpha_n)$ for 
$\alpha_n > \alpha_c$ [see Eq.(\ref{nonuni})].}
\label{fig4}
\end{figure}

In Fig. \ref{fig4} we report the obtained values of the critical exponent $t$
for $\xi/\Phi=0.2$ and for two different values of the ratio
$L/\Phi$ between the length channel and sphere diameter. 
Each square corresponds to a particular number $n$ of inter-sphere
tunneling junctions arranged in the channel (see the caption) which, 
from Eqs.(\ref{tunnel2},\ref{alfa}), also gives
the corresponding value of the tunneling exponent $\alpha_n$ reported in the abscissa.
As a function of $\alpha_n$, the critical exponent $t$ nicely follows Eq.(\ref{nonuni})
(solid curve) confirming that universal ($t\simeq 2.0$)  or nonuniversal ($t>2.0$)
behavior is obtained just by changing the number of spheres accommodated in the channels.
In our least square fittings to Eq.(\ref{scal}) we have found that the minimum $\chi^2$
is obtained by setting $b\neq 0$ and $\omega\sim 1.0$ for 
$\alpha_n < \alpha_c $ and $b=0$ for $\alpha_n>\alpha_c$. It is worth noticing that
our Monte Carlo results on the cubic lattice agrees with Eq.(\ref{nonuni}) much
better than the corresponding problem, Eqs.(\ref{distri1},\ref{distri2}), 
on the two-dimensional square lattice.\cite{lobb2,grima}

\begin{figure}[t]
\protect
\centerline{\epsfig{figure=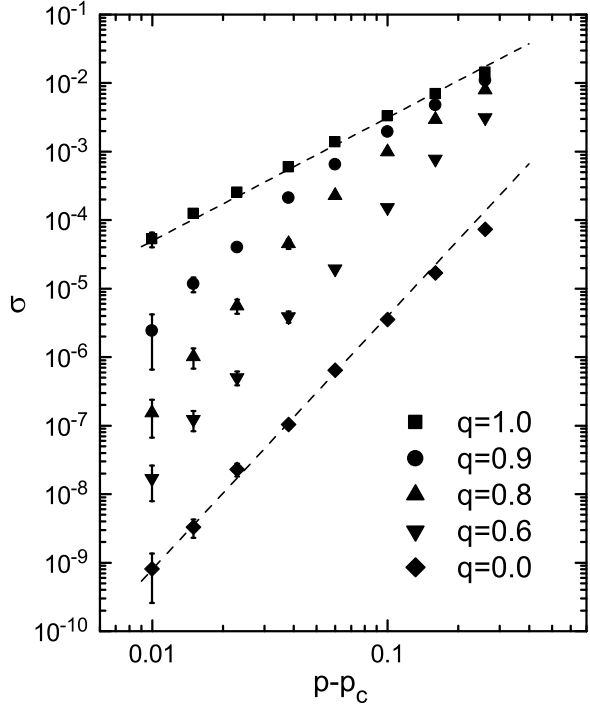,width=22pc,clip=}}
\caption{Conductivity $\sigma$ of a $40\times 40\times 40$ cubic lattice
with the bimodal distribution of Eq.(\ref{bimodal}) with $n1=9$, $n2=6$,
$L/\Phi=10$ and $\xi/\Phi=0.2$. Symbols are mean values of 10 different runs
with standard deviations given by the error bars. 
Dashed lines are fits to Eq.(\ref{eq1}) with $t=1.8\pm 0.1$ for
$q=1$ and $t=3.7\pm 0.2$ for $q=0$.}
\label{fig5}
\end{figure}

We have applied the transfer-matrix method also to the bimodal distribution of
Eq.(\ref{bimodal}) with $L/\Phi=10$, $\xi/\Phi=0.2$, $n1=9$ and $n2=6$ and 
have found that, as expected, at $p_c$ the critical exponent is nonuniversal 
already for $q=0.9$.
However what is interesting in the bimodal case is the behavior of the conductivity
$\sigma$ away from the critical thresold. The high-structured shape of $h(g)$ for 
$0 < q < 1$ reported in Fig.\ref{fig3} in fact suggests that the $p$ dependence 
of $\sigma$ could be affected by the competition between the two exponents
$\alpha_{n1} < \alpha_c$ and $\alpha_{n2} > \alpha_c$. To study this problem,
the application of the transfer-matrix method for values of the occupied channel 
concentration $p$ away from the critical thresold $p_c$ is not efficient since
the computational time of the algorithm increases as $p$ is 
moved from $p_c$.\cite{derrida1,derrida2}
Hence we have approached the problem by solving the RRN by the conjugate gradient method
which is more efficient away from the critical point.\cite{batrouni} 
The resulting $\sigma$ is
reported in Fig. \ref{fig5} for a cubic lattice of $40\times 40\times 40$ sites
and periodic boundary conditions applied to the sides not connected with the 
external potential drop. We have considered the bimodal case defined by  
$L/\Phi=10$, $n1=9$, $n2=6$ and different values of $q$. 
For $q=1$ (filled squares in fig.\ref{fig5}) all the occupied channels have $n1=9$
tunneling junctions and the conductivity is well approximated 
by Eq.(\ref{eq1}) with critical exponent $t=1.8\pm 0.1$. This value is slightly less
than the universal result $t=t_0\simeq 2.0$ and this difference signals the limitation
of extracting critical exponents from the $p$ dependence of $\sigma$ in finite size samples.
However $\sigma$ follows the power-law Eq.(\ref{eq1}) in the interval $p-p_c < 0.1-0.2$.
A nice power-law is found also for $q=0$ (filled diamonds) for which the occupied channels
have $n2=6$ number of junctions. In this case however the exponent is $t=3.7\pm 0.2$,
{\it i. e.}, slightly less than the nonuniversal value $t\simeq 4.0$ obtained 
by the transfer-matrix method (see Fig.\ref{fig4}). 
 
In Fig.\ref{fig5} we report also $\sigma$ calculated for
intermediate values of $q$. For $q=0.9$ and $q=0.8$ the $p-p_c$
dependence of the conductivity 
can be reasonably fitted by a simple power-law
only for $p-p_c < \sim 0.05$ for which we have found $t=3.2\pm 0.3$
and $t=4.1\pm 0.4$, respectively. We interpret this shrinking of 
the critical region as being due to the large contribution of the
fraction of channels with $n1=9$ to the occupied channels 
distribution function $h(g)$. However for $q=0.6$ the critical region
is already full restored and $\sigma$ follows a power-law with $t=3.8\pm 0.5$
for $p-p_c \leq 0.1$.

\section{discussion and conclusions}
\label{concl}

As shown in the previous sections, the interesting characteristic of
our segregated tunneling-percolation model is the possibility of
having universal or nonuniversal behavior of transport within the same
theoretical framework. As we have discussed, if the microscopic physical
and geometric parameters (insulating cube size $L$, sphere diameter $\Phi$, 
number of spheres and localization length $\xi$) are such that the tunneling 
factor $\alpha_n$ is larger than the critical value $\alpha_c\simeq 0.107$ then
the critical exponent $t$ is nonuniversal and follows $t=\nu+1/(1-\alpha_n)$,
otherwise transport is universal and the critical exponent is $t=t_0\simeq 2.0$.
This universal/nonuniversal crossover is experimentally observed in thick-film
resistors for which is reported to vary between $t\simeq 2.0$ and
$t\simeq 5.0$ also for mixtures of chemically identical constituents.
It can be argued that different fabrication procedures (for example
firing temperature) affects the microstructure leading to different effective
values of $\alpha_n$. Of course our model is oversimplified in the sense that
interactions between the conducting and insulating phases are completely
neglected. However it is remarkable that only two assumptions, quasi-one 
dimensionality of the conducting channels and Poisson distribution of the
position of the spheres inside the channels, are sufficient to give rise to
such rich phenomenology.

The model discussed in this paper captures the essential physics, but eventually
it can be further generalized to include more realistic features. 
For example, it is possible to account for
different sizes of the conducting spheres in a straightforward manner, since 
also for this case the one-dimension nearest-neighbour distance distribution 
function is provided by an analytical and exact expression.\cite{lu} Also the tunneling 
expression Eq.(\ref{tunnel1}) can be refined by including, for example,
charging energies or distribution functions for the tunneling factor $\xi$.

\acknowledgments
We are grateful to Isaac Balberg for interesting discussions.
This work is part of TOPNANO 21 project n.5557.2.

\end{document}